\documentclass{article}

\usepackage{PRIMEarxiv}

\usepackage[utf8]{inputenc} 
\usepackage[T1]{fontenc}    
\usepackage{hyperref}       
\usepackage{url}            
\usepackage{booktabs}       
\usepackage{amsfonts}       
\usepackage{nicefrac}       
\usepackage{microtype}      
\usepackage{lipsum}
\usepackage{fancyhdr}       
\usepackage{graphicx}       
\graphicspath{{media/}}     
\usepackage{subcaption}

\title{PolicyBot - Reliable Question Answering over Policy Documents
}

\author{
    Gautam Nagarajan \\
    School of Engineering  \\
    Shiv Nadar University \\
    Noida, India \\
    \texttt{gautam.nraj@gmail.com} \\
    \And
  Omir Kumar \\
  Centre for Responsible AI \\
  IIT Madras \\
  Chennai, India\\
  \texttt{omir@cerai.in} \\
   \And
  Sudarsun Santhiappan \\
  Centre for Responsible AI \\
  Wadhwani School of Data Science \& AI\\
  IIT Madras \\
  Chennai, India\\
  \texttt{sudarsun@dsai.iitm.ac.in} \\
}

\begin{document}
\maketitle

\begin{abstract}
All citizens of a country are affected by the laws and policies introduced by their government. These laws and policies serve essential functions for citizens. Such as granting them certain rights or imposing specific obligations. However, these documents are often lengthy, complex, and difficult to navigate, making it challenging for citizens to locate and understand relevant information. This work presents \texttt{PolicyBot}, a retrieval-augmented generation (RAG) system designed to answer user queries over policy documents with a focus on transparency and reproducibility. The system combines domain-specific semantic chunking, multilingual dense embeddings, multi-stage retrieval with reranking, and source-aware generation to provide responses grounded in the original documents. We implemented citation tracing to reduce hallucinations and improve user trust, and evaluated alternative retrieval and generation configurations to identify effective design choices. The end-to-end pipeline is built entirely with open-source tools, enabling easy adaptation to other domains requiring document-grounded question answering. This work highlights design considerations, practical challenges, and lessons learned in deploying trustworthy RAG systems for governance-related contexts.
\end{abstract}


\section{Introduction}

Public policy documents form the backbone of governance, defining laws, regulations, entitlements, and procedures that affect citizens, institutions, and industries. However, these documents are often dense, verbose, lengthy, and inaccessible to non-experts. Legal jargon, bureaucratic phrasing, and intricate cross-referencing create significant barriers to understanding, even for professionals familiar with the policy domain. As a result, individuals seeking specific information to make decisions, verify claims, or understand their rights often rely on secondhand summaries, expert intermediaries, or extensive manual reading. These approaches can be slow, expensive, error-prone, and prone to omissions. Therefore, it may be argued that laws and policies are often inaccessible to the very people (citizens) for whom they are made.  

Existing question answering (QA) and retrieval-augmented generation (RAG) systems have shown promise in domains such as general web search, customer support, and scientific literature. However, they face particular challenges in the legal and policy document domain. Conventional methods often struggle with the length and complexity of these documents, resulting in incomplete retrieval, loss of context, or factual errors in the generated responses. Moreover, many systems lack reliable citation tracing, making it difficult for users to verify information—a critical need in governance and legal contexts. Hallucination, where systems fabricate content not present in the source, further undermines trust in automated policy QA tools.

To address these challenges, we present \textbf{PolicyBot}, a retrieval-augmented generation system designed specifically for policy document question answering. \texttt{PolicyBot} integrates domain-specific semantic chunking with a multi-stage retrieval pipeline that includes multilingual dense embeddings, RAG-Fusion, Hypothetical Document Embeddings (HyDE), and reranking via a cross-encoder model. It employs source-aware generation and citation tracing to ensure factual accuracy and reduce hallucinations. The system runs efficiently on consumer-grade hardware, utilizing the \texttt{Gemma~3n}\footnote{https://deepmind.google/models/gemma/gemma-3n/} language model via \texttt{Ollama}\footnote{https://github.com/ollama/ollama}, which enables deployment in low-bandwidth or offline settings. It supports context-aware, multi-turn conversations by preserving relevant query history and maintaining coherence over extended interactions.

The key contributions of this work are:
\begin{itemize}
  \item \textbf{Factual accuracy and traceability:} A multi-layered hallucination control strategy combining document-grounded generation, direct quotations, and explicit source chunk display.
  \item \textbf{Advanced retrieval pipeline:} Domain-specific semantic chunking and multi-stage retrieval with HyDE, RAG-Fusion, and reranking to improve relevance and handle ambiguous queries.
  \item \textbf{Hardware-efficient local deployment:} An open-source, lightweight LLM architecture enabling private, low-latency inference on consumer-grade hardware.
  \item \textbf{Context-aware multi-turn interaction:} Support for coherent, multi-turn dialogue grounded in prior conversation history.
\end{itemize}

\texttt{PolicyBot} is designed to benefit diverse user groups, including students seeking clarity on laws, policies, and regulations; journalists fact-checking policy claims; NGOs and citizens verifying their rights; public sector professionals reviewing operational rules; and researchers examining law and policy. Its offline capability makes it particularly relevant for under-resourced communities, enhancing equitable access to reliable policy information. The \texttt{PolicyBot} not only helps improve accessibility to applicable laws and policies for citizens, but also ensures that respective governments and related institutions can reach a greater number of people. 

\textbf{Paper Outline.} The remainder of this paper is organized as follows. 
Section~\ref{sec:system} describes the system architecture and design choices in detail. 
Section~\ref{sec:setup} presents the experimental setup, including datasets, metrics, and tested configurations. 
Section~\ref{sec:deployment} discusses deployment considerations and key lessons learned. 
Section~\ref{sec:ethics} addresses ethical considerations relevant to policy document QA. 
Section~\ref{sec:future} outlines potential extensions to enhance scalability, multilingual support, and user-centered evaluation
Section~\ref{sec:conclusion} concludes by reflecting on how the proposed \texttt{PolicyBot} system advances reliable, transparent, and accessible question answering over policy documents.
Section~\ref{sec:demo_source} provides the links to the live demo and the source code of the proposed \texttt{PolicyBot} system.





\begin{figure}[t]
    \centering
    \includegraphics[scale=0.3]{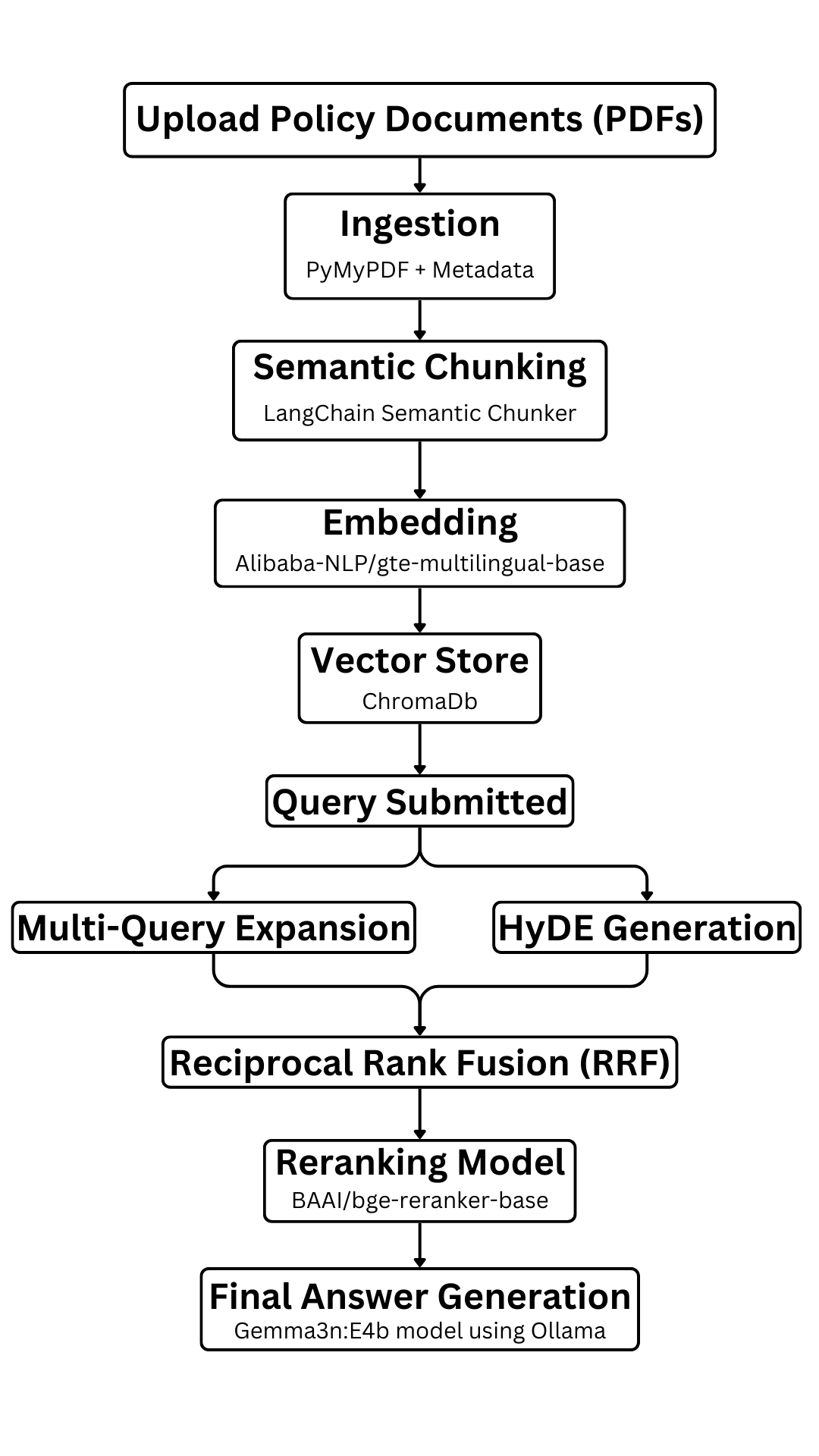}  
    \caption{Architecture of the \texttt{PolicyBot} system.}
    \label{fig:architecture}
\end{figure}

\section{System Design and Architecture}
\label{sec:system}

\texttt{PolicyBot} is implemented as a modular retrieval-augmented generation (RAG) pipeline optimized for the characteristics of public policy documents. The architecture comprises seven main components: data ingestion, semantic chunking, embedding and indexing, retrieval and reranking, generation, citation, and hallucination control, as well as the user interface. Figure~\ref{fig:architecture} illustrates the overall workflow.

\subsection{Data Ingestion}
The current implementation only supports PDF files as input. The \texttt{PyMuPDF}\footnote{https://pymupdf.readthedocs.io/en/latest/} library is used for fast and accurate text extraction. Each page of the input PDF is treated as a separate document, and its text content is stripped of leading and trailing whitespace before further processing. For each page, metadata is stored that includes the page number and the original file name. No additional preprocessing, such as header or footer removal or language detection, is performed, as the text extraction capabilities of \texttt{PyMuPDF} are sufficient for the target documents.

\subsection{Semantic Chunking}
To segment documents into semantically coherent units, \texttt{PolicyBot} utilizes the experimental semantic text splitter \cite{langchain_semantic_chunker} of  LangChain\footnote{https://python.langchain.com/docs/introduction/}. This splitter segments the text based on logical and semantic boundaries rather than fixed-size windows. The segmentation process uses \texttt{standard\_deviation} as the \texttt{breakpoint\_threshold\_type} and sets the value of \texttt{breakpoint\_threshold\_amount} to \texttt{1.0}, enabling the chunking algorithm to adapt to the natural structure of the policy documents. This approach preserves the integrity of legal clauses and section-level semantics, reducing the likelihood of losing critical contextual dependencies. Empirically, this method consistently outperformed alternatives, such as fixed-length or recursive chunking, in the domain of policy documents.

\subsection{Embedding and Indexing}
We use \texttt{alibaba/gte-multilingual-base} \cite{zhang2024mgte} to create the rich embeddings of the semantic chunks.  We chose the model for its strong performance across multilingual factual question-answering tasks in the SciFact dataset \cite{Wadden2020FactOF} from the BEIR benchmarking framework \cite{thakur2021beir}. While the model supports multiple languages, the current deployment primarily focuses on policy documents in English.

We use ChromaDB \cite{chroma_db} to store the vector embeddings locally. ChromaDB is configured with the default HNSW (Hierarchical Navigable Small World) \cite{8594636} vector index for efficient approximate nearest neighbor (ANN) search. Cosine similarity is used as the distance metric for retrieval. This combination of dense embeddings and ANN indexing provides a balance of semantic fidelity and query-time efficiency, making it suitable for interactive use.

\subsection{Retrieval and Reranking}
\texttt{PolicyBot} employs a multi-stage retrieval pipeline to optimize recall and relevance before generating answers.

\paragraph{HyDE Retrieval} \cite{gao2022precisezeroshotdenseretrieval}: In the first stage, the \texttt{Gemma~3n} model is prompted with a carefully crafted system prompt, the user's query, and a summary of the entire document to generate a \emph{hypothetical answer} that might appear in the policy text. The embedding of this hypothetical answer is then used to retrieve a set of candidate chunks from ChromaDB.

\paragraph{Multi-query Generation} \cite{tahata2023diagonalsparametersymmetrymodelproperty}: The system generates five semantically diverse re-wordings of the user's original query, again using the document summary for context. Each reworded query is embedded and used to retrieve additional candidate chunks.

\paragraph{Reciprocal Rank Fusion (RRF)} \cite{10.1145/1571941.1572114}: Candidate sets from the HyDE and multi-query retrieval stages are merged using Reciprocal Rank Fusion. RRF assigns a score to each chunk based on its rank across the different retrieval lists, prioritizing chunks that consistently appear near the top in multiple rankings. Top-$p$ filtering is applied to the fused results to select a dynamically determined set of highly relevant chunks.

\paragraph{Reranking:} The final candidate set from RRF is passed to the \texttt{BAAI/bge-reranker-base} \cite{bge_embedding} cross-encoder model, which assigns a fine-grained relevance score for each chunk concerning the original user query. A second top-$p$ filtering step selects the highest-confidence chunks for downstream generation.

\subsection{Generation}
Answer generation is performed by the \texttt{Gemma~3n} model running locally via \texttt{Ollama}. The model is configured with a context length of 32{,}000 tokens, although the full limit is rarely reached. The temperature is set to 0.1 to encourage deterministic, factual outputs over creative variation. No explicit maximum output length is imposed. The generation process is orchestrated by a custom LangChain pipeline using a domain-specific system prompt that enforces several constraints:
\begin{itemize}
    \item Only information from retrieved chunks should be used to generate the answer.
    \item Direct quotations from the source are preferred wherever applicable.
    \item If sufficient supporting context is absent, the model must respond with ``not enough context'' rather than attempting to answer.
\end{itemize}
This prompt design serves as a central safeguard against hallucination, ensuring that all generated outputs are grounded in the policy document.

\subsection{Citation and Hallucination Control}
\texttt{PolicyBot} employs a dual approach to ensure factual accuracy and maintain traceability. First, hallucination control is enforced through the generation prompt described above, which explicitly restricts the model from introducing knowledge that's outside the supplied Policy documents. Second, citations are provided directly to the user via the front-end. Beneath each answer, a collapsible section displays the exact text chunks that were passed to the generation model, along with their metadata. This allows users to verify every statement against the original source. Dynamic top-$p$ filtering during retrieval further reduces noise in the input context, thereby minimizing the likelihood of unsupported content.

\subsection{User Interface}
We implement an indicative user interface using \texttt{Streamlit}\footnote{https://docs.streamlit.io/}, providing an interactive and accessible front end. The interface supports free-form queries, displays multi-turn conversation history, and embeds interactive citations that expand on demand. This design is intended to serve diverse audiences, including policy officers, journalists, NGOs, and citizens, without requiring technical expertise. While the current deployment is local, the architecture allows straightforward adaptation to cloud or on-premises environments.

\section{Experimental Setup}
\label{sec:setup}

To evaluate and select the optimal components for the retrieval-augmented generation (RAG) pipeline, a series of controlled experiments was conducted. The evaluation focused on three key components: the large language model (LLM) for answer generation, the embedding model for retrieval, and the chunking strategy for document segmentation.

\subsection{LLM Decision}
The large language model (LLM) is the central component of the \texttt{PolicyBot}'s answer generation pipeline, responsible for producing responses that are factually correct and contextually relevant to user queries. For this application, the LLM needed to satisfy three crucial requirements:
\begin{enumerate}
    \item Minimize hallucinations and ensure that all generated content is grounded in the retrieved context.
    \item Maintain a professional and neutral tone suitable for interpreting policy-related queries.
    \item Operate efficiently on consumer-grade hardware without significant degradation in quality.
\end{enumerate}

To evaluate candidate models, a custom benchmark dataset was created consisting of policy-related questions, their associated source document excerpts, and reference answers. This dataset was specifically curated to reflect the linguistic and structural complexity of real-world policy documents, including long clauses, conditional statements, and nested definitions. Six open-source LLMs available via the \texttt{Ollama} framework were selected for testing:
\begin{itemize}
    \item \texttt{deepseek-r1:8b} \cite{deepseekr1_8b}
    \item \texttt{mistral:7b} \cite{mistral7b_instruct_v02}
    \item \texttt{llama3.1:8b} \cite{llama3.1-8b-ollama}
    \item \texttt{qwen2.5:7b} \cite{qwen2.5}
    \item \texttt{gemma:7b} \cite{gemma_2024}
    \item \texttt{gemma 3n:e4b} \cite{gemma_3n_2025}
\end{itemize}

Each model was evaluated using the same retrieval context and identical system prompt. The prompt was designed to explicitly restrict the model to using only the provided context, encourage the inclusion of direct quotations where relevant, and instruct the model to respond with ``not enough context'' if it could not confidently answer the question. This ensured that the primary difference in performance would stem from the model's inherent capabilities rather than variations in retrieval quality or prompt structure. The evaluation procedure involves the following steps:
\begin{itemize}
    \item Running each model locally on identical hardware to ensure a fair comparison in latency and resource usage.
    \item Assessing each generated answer on three primary axes: \textit{factual accuracy} (measured against the reference answer), \textit{adherence to the system prompt} (compliance with hallucination control and citation requirements), and \textit{stylistic quality} (professional tone, clarity, and neutrality).
    \item Computing a set of quantitative metrics for each model: similarity, ROUGE-L \cite{DBLP:journals/jocs/CitarellaBCMBT25}, BLEU, METEOR, BERT-based precision, recall, and $F_1$ scores \cite{DBLP:journals/corr/abs-2109-14250}. These metrics capture both lexical overlap and semantic similarity between generated answers and reference answers.
    \item Recording the proportion of cases in which the model explicitly responded with ``not enough context,'' serving as an indicator of conservative generation behavior where relevant information wasn't available in the retrieved context.
\end{itemize}
Across these metrics, \texttt{gemma 3n:e4b} consistently achieved the highest performance, demonstrating a strong balance between factual accuracy, semantic fidelity, and stylistic appropriateness, while adhering closely to the constraints on hallucination mitigation.  From this evaluation, \texttt{gemma 3n:e4b} emerged as the most balanced choice. It consistently provided accurate, well-grounded responses, demonstrated strong adherence to hallucination mitigation instructions, and maintained a clear, professional tone. Larger models such as \texttt{llama3.3:8b} and \texttt{qwen2.5:7b} occasionally generated more verbose answers, but these often contained unsupported statements. Importantly, \texttt{gemma 3n:e4b} achieved this performance while running efficiently, making it well-suited for the local, resource-constrained deployment environments targeted by \texttt{PolicyBot}.

\subsection{Embedding Model Selection}
High-quality dense embeddings are critical to the retrieval stage of any RAG system: they determine whether semantically relevant document chunks are surfaced for downstream reranking and generation. For \texttt{PolicyBot}, the embedding model needed to satisfy several practical and domain-specific requirements:
\begin{enumerate}
    \item \textbf{Semantic fidelity:} The embeddings must capture subtle semantic relationships in policy text (e.g., conditionals, exceptions, references) so that relevant clauses are retrieved even when queries are phrased differently from the source wording.
    \item \textbf{Robustness to domain wording:} The model should handle formal, legalistic, and bureaucratic language common in policy documents.
    \item \textbf{Multilingual support:} While the current system primarily handles English documents, multilingual capability was desirable for potential future extensions.
    \item \textbf{Retrieval efficiency and compatibility:} The embeddings should work well with approximate nearest neighbor (ANN) indexing (ChromaDB with HNSW) and enable fast interactive queries.
\end{enumerate}
\begin{figure}[htbp]
  \centering
  \begin{subfigure}[b]{\linewidth}
  \centering
    \includegraphics[scale=0.4]{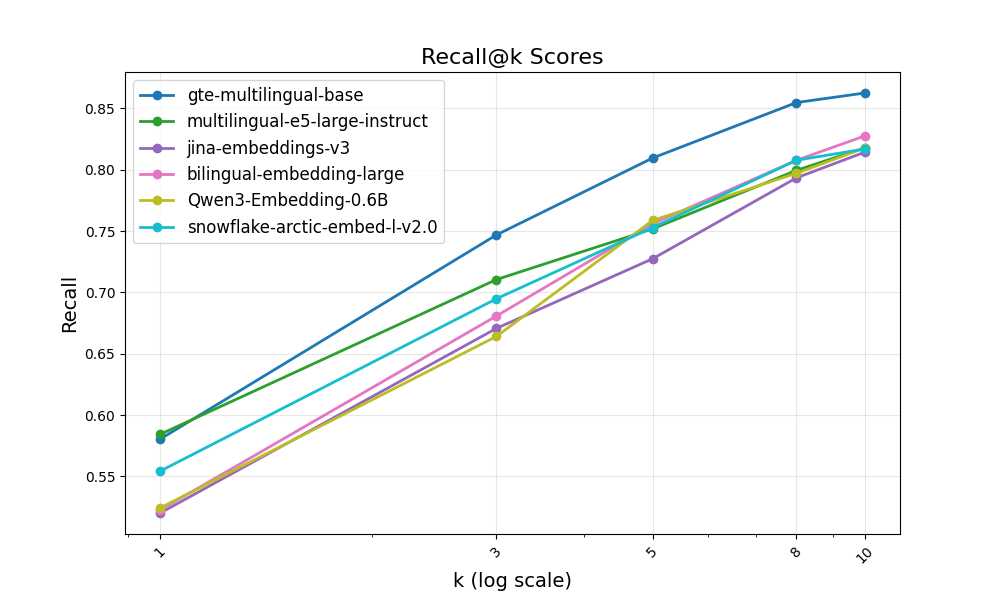}
    \caption{Recall scores for different embedding models}
    \label{fig:3a}
  \end{subfigure}
  \begin{subfigure}[b]{\linewidth}
  \centering
    \includegraphics[scale=0.4]{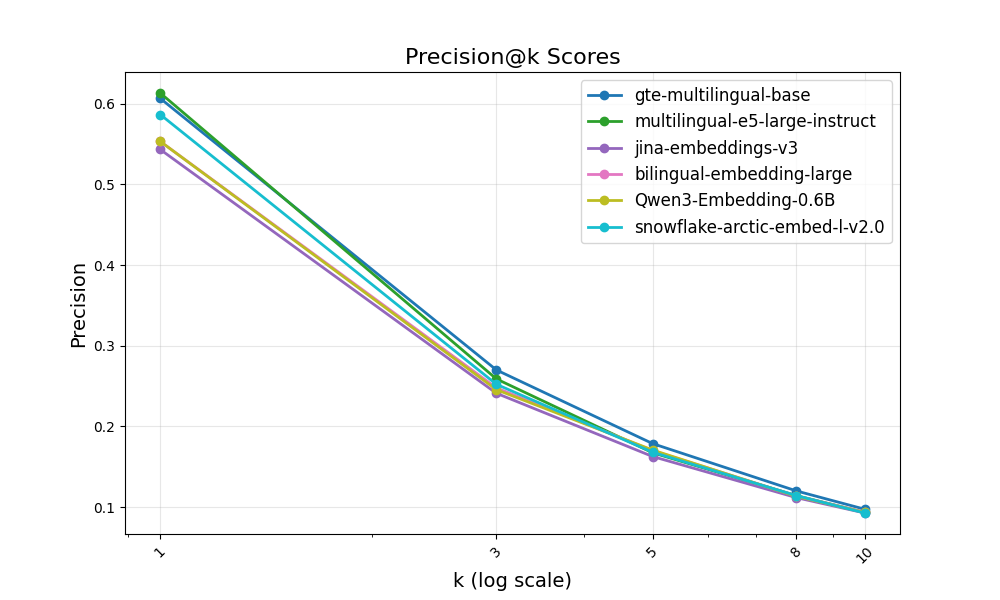}
    \caption{Precision scores for different embedding models}
    \label{fig:3b}
  \end{subfigure}
  \begin{subfigure}[b]{\linewidth}
  \centering
    \includegraphics[scale=0.4]{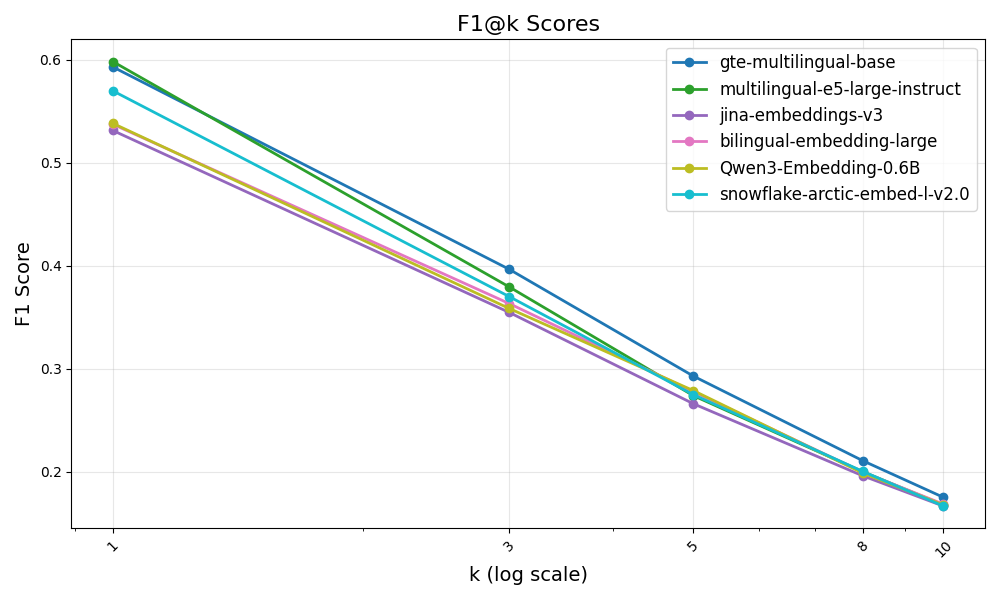}
    \caption{$F_1$ scores for different embedding models}
    \label{fig:3c}
  \end{subfigure}

  \caption{Comparative performance of embedding models at various retrieval depths}
  \label{fig:embedding-benchmark}
\end{figure}

To perform a standardized and reproducible comparison, we used the BEIR evaluation framework \cite{thakur2021beir} with the SciFact dataset \cite{Wadden2020FactOF}. SciFact was chosen because it provides a factual QA-style retrieval task that stresses semantic matching between queries and scientific claims—an appropriate proxy for the factual, clause-oriented matching required in policy documents. Retrieval performance was evaluated at multiple retrieval depths ($k$), and models were compared using precision, recall, and $F_1$ score at these depths to quantify both exact-match and soft-match retrieval quality.  The candidate embedding models are listed below.

\begin{itemize}
    \item \texttt{Alibaba-NLP/gte-multilingual-base} \cite{zhang2024mgte}
    \item \texttt{intfloat/multilingual-e5-large-instruct} \cite{wang2024multilingual}
    \item \texttt{jinaai/jina-embeddings-v3} \cite{sturua2024jinaembeddingsv3multilingualembeddingstask}
    \item \texttt{Lajavaness/bilingual-embedding-large} \cite{conneau2019unsupervised, reimers2019sentence,thakur2020augmented}
    \item \texttt{Qwen/Qwen3-Embedding-0.6B} \cite{qwen3embedding}
    \item \texttt{Snowflake/snowflake-arctic-embed-l-v2.0} \cite{snowflake_arctic_embed_l_v2}
\end{itemize}

The evaluation procedure consists of the following steps:
\begin{itemize}
    \item \textbf{Index construction:} For each candidate model, embeddings were generated for the SciFact corpus and indexed in a local ChromaDB instance configured with the default HNSW ANN index and cosine similarity as the distance metric.
    \item \textbf{Retrieval experiments:} For each model, standard BEIR retrieval scripts were used to run queries against the index, collecting retrieval results at several $k$ values. 
    \item \textbf{Metric computation:} Precision, recall, and $F_1$ score were computed at each $k$ to assess how reliably each model surfaced relevant documents across retrieval depths. These metrics capture the trade-off between precision at shallow depths and recall at deeper retrievals.
    \item \textbf{Controlled conditions:} All experiments were executed on the same hardware and indexing configurations to ensure comparability across models. Embedding generation was batched to reflect realistic production batching behavior, and caching was consistently used across runs to minimize measurement variance resulting from I/O differences.
\end{itemize}

Figure~\ref{fig:embedding-benchmark} summarizes the comparative retrieval performance across the tested models. \texttt{Alibaba-NLP/gte-multilingual-base} consistently achieved the highest $F_1$ scores across a range of retrieval depths, indicating a favorable balance between precision and recall for this task. In particular, \texttt{gte-multilingual-base} exhibited stronger performance at moderate depths (e.g., $k=5$ and $k=10$), which are commonly used in RAG pipelines to assemble candidate contexts for reranking and generation.

Beyond raw retrieval metrics, practical considerations also informed the selection. \texttt{gte-multilingual-base} demonstrated stable behavior with ChromaDB's HNSW index and produced embeddings that integrated efficiently into our multi-stage retrieval pipeline (HyDE + multi-query + RRF + reranker). Its multilingual capability provides flexibility for future extensions without compromising current English-language performance. Based on these quantitative and qualitative observations, \textsf{Alibaba-NLP/gte-multilingual-base} was selected as the embedding model for \texttt{PolicyBot}'s production pipeline.

\subsection{Chunking Strategies}
The chunking strategy directly determines the semantic coherence of document chunks and, consequently, the quality of retrieval in a RAG system. For the \texttt{PolicyBot}, chunking is especially critical because policy documents often contain long, legally precise clauses, where splitting at arbitrary points can distort meaning or omit essential context.

\subsubsection{Dataset Preparation and Modification}
To benchmark different chunking strategies, we use the Microsoft/Wiki\_QA dataset~\cite{yang-etal-2015-wikiqa} as the base. This dataset comprises pairs of natural language questions and corresponding sentences from Wikipedia articles, with each sentence provided separately. For this evaluation, the dataset was modified as follows:
\begin{enumerate}
    \item All sentences belonging to the same Wikipedia article title were concatenated into a single paragraph, preceded by the article title as a heading. This was followed immediately by the next article title and its corresponding concatenated content, producing a continuous, multi-article corpus.
    \item The resulting structure created a long, uninterrupted text stream suitable for testing various chunking mechanisms under realistic retrieval conditions.
    \item The original question–answer mappings were preserved at the sentence level, allowing precise verification of whether a retrieved chunk contained the gold-standard answer sentence.
\end{enumerate}

\begin{figure}[htbp]
    \centering
    \begin{subfigure}[b]{\linewidth}
        \centering
        \includegraphics[scale=0.4]{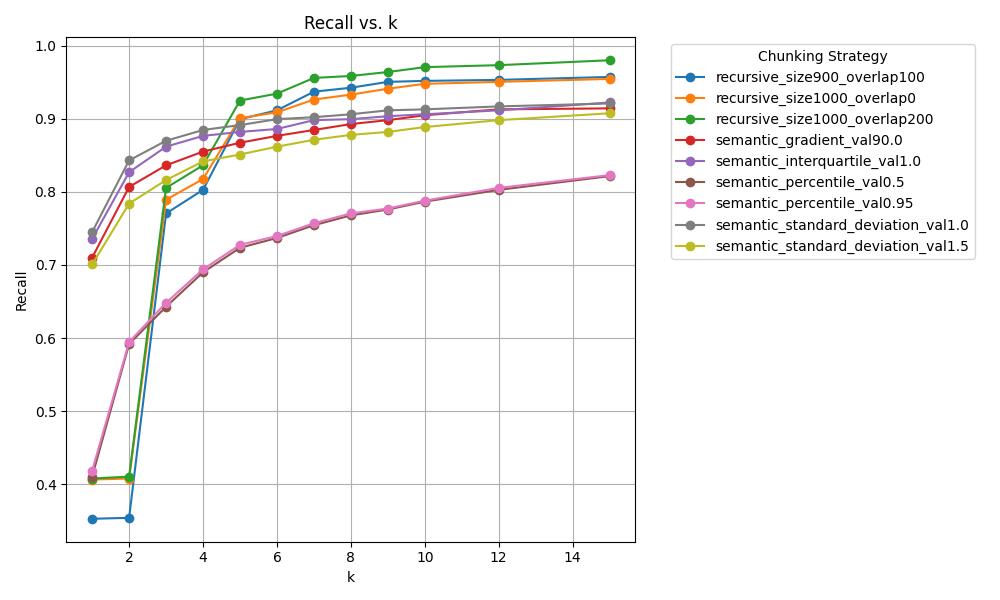}
        \caption{Recall scores for different chunking strategies}
        \label{fig:chunking-recall}
    \end{subfigure}

    \vspace{1em}

    \begin{subfigure}[b]{\linewidth}
        \centering
        \includegraphics[scale=0.4]{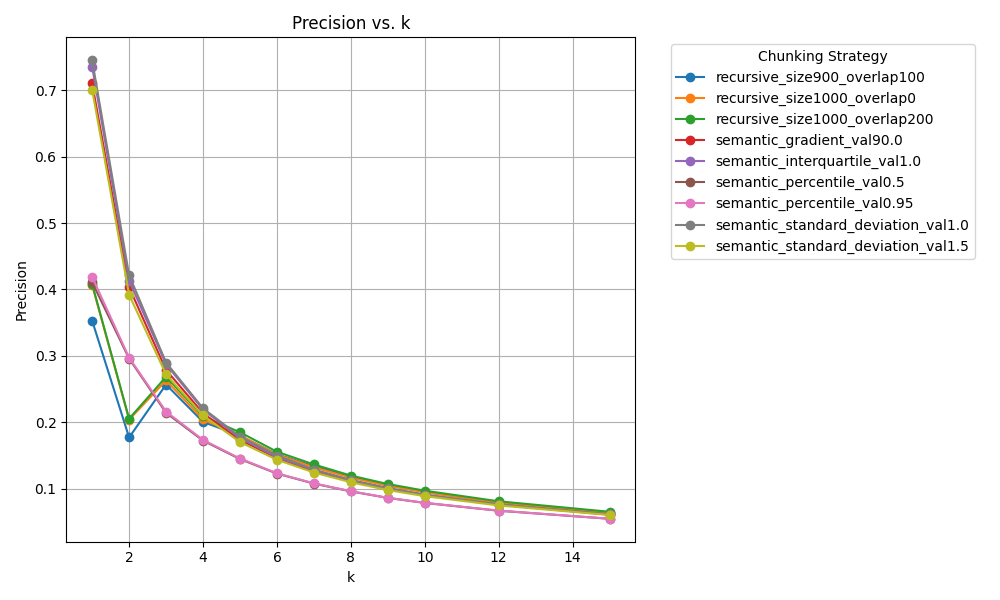}
        \caption{Precision scores for different chunking strategies}
        \label{fig:chunking-precision}
    \end{subfigure}

    \vspace{1em}

    \begin{subfigure}[b]{\linewidth}
        \centering
        \includegraphics[scale=0.4]{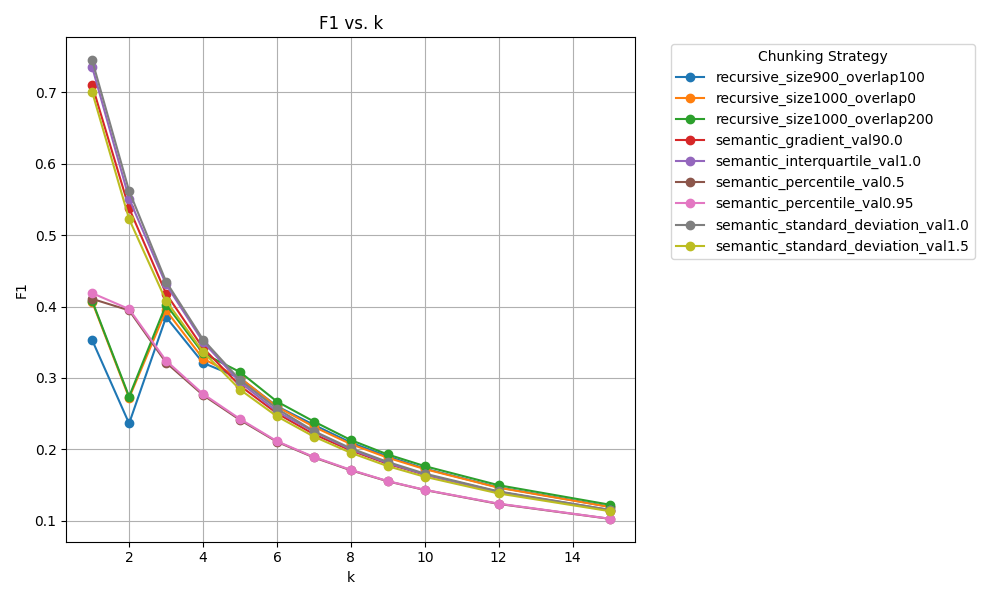}
        \caption{$F_1$ scores for different chunking strategies}
        \label{fig:chunking-f1}
    \end{subfigure}
    
    \caption{Comparison of chunking strategies at various retrieval depths}
    \label{fig:chunking-benchmark}
\end{figure}

\subsubsection{Relevance Definition}
For evaluation purposes, a retrieved chunk was considered \emph{relevant} to a query if it contained the complete sentence that was labeled as the correct answer in the original Wiki\_QA dataset. If the answer sentence was not fully contained within the chunk, the retrieval was marked as non-relevant, even if partial information was present. This ensured strict relevance criteria aligned to return self-contained, verifiable policy clauses.

\subsubsection{Chunking Strategies Tested}
Two primary segmentation approaches, each with multiple configurations, were evaluated:
\begin{itemize}
    \item \textbf{Recursive Chunking} \cite{langchain_recursive_character_text_splitter}: Implemented using LangChain's \texttt{RecursiveCharacterTextSplitter}, this method segments text based on a fixed character length, attempting to break on natural boundaries (paragraphs, sentences) when possible, but without semantic analysis. Multiple configurations were tested by varying the chunk size (e.g., 750, 800, 1000 characters) and overlap length (200--250 characters) to assess the trade-off between retrieval coverage and precision.
    \item \textbf{Semantic Chunking} \cite{langchain_semantic_chunker}: Implemented using LangChain's experimental semantic chunker, which leverages statistical variance in sentence embeddings to determine natural breakpoints in the text. Several parameterizations were explored across different breakpoint detection methods, including gradient, percentile, and standard deviation thresholds, to evaluate their impacts on semantic coherence and retrieval performance.
\end{itemize}

\subsubsection{Evaluation Procedure}
For each chunking strategy, multiple configurations were tested to account for variations in chunk size, overlap, and sensitivity to semantic breakpoints. In the case of recursive chunking, configurations included fixed sizes of 750, 800, and 1000 characters with overlaps ranging from 200--250 characters. For semantic chunking, configurations varied across breakpoint detection methods, including gradient, percentile, and standard deviation thresholds (e.g., gradient at 0.75, percentile at 0.9, and standard deviation at 1.0). For each configuration:
\begin{enumerate}
    \item The modified dataset documents were segmented according to the chosen method and parameter set.
    \item Chunks were embedded using the selected production embedding model (\texttt{gte-multilingual-base}) and indexed in ChromaDB with an HNSW index and cosine similarity.
    \item Original Wiki\_QA questions were used as retrieval queries, returning the top-$k$ chunks for multiple values of $k$.
    \item Retrieval performance was measured using precision, recall, and $F_1$ score, where relevance was determined using the strict criterion described above.
\end{enumerate}

\subsubsection{Results}
The semantic chunking configuration consistently outperformed recursive chunking across low $k$ values. Its ability to preserve complete logical units of text—such as entire clauses or multi-sentence definitions—led to higher recall without sacrificing precision. This property is particularly valuable in the policy document setting, where losing part of a clause can alter the interpretation of the text.

\section{Deployment \& Lessons Learned}
\label{sec:deployment}

The system was deployed as a web-based application, designed to operate in both online and offline settings. It uses a \texttt{Streamlit} interface for interaction and runs entirely on local infrastructure, leveraging \texttt{Ollama} to host the \texttt{Gemma~3n} model and ChromaDB for vector storage. This self-contained design enables deployment in environments with limited or no internet connectivity, ensuring privacy of sensitive policy documents and eliminating reliance on external APIs. The deployment target was to create a tool suitable for diverse stakeholders, such as students, journalists, NGOs, and public officials.

\subsection{Key Deployment Challenges}

During deployment, several practical challenges emerged. One significant factor was the trade-off between latency and accuracy. Increasing the number of retrieved chunks or expanding the context length generally improves response accuracy but at the cost of slower inference. This necessitated careful tuning of retrieval parameters to strike a balance between relevance and response time.

A second challenge was multilingual capability. While the chosen embedding model (\texttt{gte-multilingual-base}) supports multiple languages, the current implementation was evaluated primarily on English policy documents. Extending reliable performance to other languages will require additional benchmarking.

Hardware constraints also influenced the deployment strategy. The system was tested on a consumer-grade NVIDIA GeForce GTX 1650 GPU. This limited the size of models that could be used without excessive latency or memory pressure. As a result, the architecture prioritized a lightweight LLM and an efficient retrieval pipeline.

\subsection{Lessons Learned}

Several insights were gained during the deployment process. First, incorporating direct citation tracing in the interface significantly improved the ease of cross-verifying responses. By showing the exact text chunks that supported a generated answer, the system allowed rapid validation of the output against the source document.

Second, semantic chunking proved critical for this domain. Maintaining the logical coherence of policy clauses within a chunk substantially improved retrieval quality compared to fixed-size segmentation.

Finally, local inference was found to be a viable strategy for policy document QA. By carefully selecting and optimizing models and retrieval parameters, it was possible to deliver grounded, context-aware responses with minimal infrastructure requirements.

\section{Ethical Considerations}
\label{sec:ethics}

The deployment of a retrieval-augmented generation (RAG) chatbot for policy documents raises several critical ethical considerations. Given that the system is intended for contexts where the accuracy and transparency of information have direct social and governance implications, it is crucial to address potential risks in bias, fairness, transparency, and privacy.

\subsection{Bias and Fairness}

Large Language Models (LLMs) can inherit biases from their training data, which may manifest in subtle ways when interpreting or summarizing policy documents. Such biases could lead to the prioritization of specific interpretations over others or the omission of relevant but less frequently represented perspectives. This is particularly significant in policy contexts, where nuanced phrasing can alter the perceived meaning of a clause. While the current system minimizes this risk by constraining the model to operate strictly within the retrieved document context, residual bias may still arise through the ranking and selection of chunks. Future work should explore bias detection frameworks and the integration of fairness-aware retrieval techniques to ensure balanced representation in retrieved content.

\subsection{Transparency in Governance Contexts}

Transparency is essential for maintaining accountability in policy interpretation. The chatbot addresses this by implementing a citation tracing mechanism, wherein the exact text chunks used for answer generation are displayed alongside the output. This design enables rapid cross-verification of responses against the source material, reducing the likelihood of unsubstantiated claims. In governance contexts, where multiple stakeholders may scrutinize decisions, the ability to trace an answer to its originating source is a safeguard against misinformation and misinterpretation. However, transparency also depends on the completeness of retrieved content; if the retrieval pipeline fails to include key relevant passages, even a fully cited answer may be incomplete.

\subsection{Privacy and Data Protection}

Policy documents can range from publicly available legislation to internal regulatory guidelines that may contain sensitive or confidential information. The local-first design of the system, with no external API calls during query processing, mitigates the risk of sensitive data leakage. All document parsing, embedding, and inference occur within the local deployment environment, ensuring that the content never leaves the host machine. Nevertheless, when used in shared or institutional settings, additional safeguards—such as encrypted storage of embeddings and access control for the user interface—should be considered to prevent unauthorized access. Furthermore, if future versions incorporate multilingual or cross-border policy analysis, varying data protection regulations (e.g., GDPR, DPDP Act) will need to be explicitly addressed in the system design.

\subsection{Ethical Deployment Practices}

Ethical deployment of such systems requires an understanding that they are assistive tools, not authoritative sources of legal interpretation. Outputs should be positioned as aids to human decision-making, rather than definitive rulings. Disclaimers in the interface can reinforce this distinction, and careful user training can further mitigate risks of over-reliance. Additionally, reproducibility and auditability of system outputs—achieved through open-source implementation and fixed-version component releases—help ensure that results can be independently verified.

In summary, while the current architecture incorporates mechanisms for transparency and privacy, continuous attention to bias mitigation, data governance, and the framing of outputs is necessary for the responsible deployment of policy contexts. These considerations are not static; they must evolve in tandem with advances in LLM capabilities, retrieval methods, and the legal frameworks governing digital policy tools.

\section{Future Work}
\label{sec:future}

While the current system demonstrates the feasibility and benefits of a domain-specific Retrieval-Augmented Generation (RAG) pipeline for policy documents, several promising avenues for further development remain. These directions span technical, linguistic, and user-centered aspects to expand applicability and impact.

\subsection{Domain Adaptation to Broader Legal and Regulatory Texts}
The present work focuses on public policy documents; however, the architecture can be extended to handle diverse legal corpora, including statutes, judicial opinions, regulatory guidelines, and international treaties. Each of these domains presents unique structural and linguistic challenges: statutes often have deeply nested numbering schemes, judicial opinions contain extensive citations and precedent references, and treaties incorporate multilingual parallel clauses. Domain adaptation will require refining chunking strategies to preserve logical integrity in these formats, as well as integrating embeddings fine-tuned on legal corpora to improve retrieval accuracy.

\subsection{Multilingual and Cross-Jurisdictional Support}
Although the selected embedding model supports multilingual input, the current deployment primarily operates in English. Expanding to fully multilingual policy datasets would make the system valuable in multilingual governance contexts—such as Indian state government portals, EU policy repositories, or UN reports. This extension would require curated multilingual corpora, rigorous cross-lingual retrieval benchmarking, and evaluation of translation fidelity when retrieved content is displayed alongside its original-language source.

\subsection{Enhanced Hallucination Control}
The current system enforces hallucination control through strict prompting, citation tracing, and top-$p$ filtering. Future enhancements could include retrieval confidence estimation (e.g., using similarity score thresholds), automated factuality checks against authoritative sources, and fine-tuning the LLM on citation-heavy legal datasets to strengthen grounded generation. Another promising approach is multi-model cross-verification, where two independent models must converge on the same answer before it is presented to the user.

\subsection{User-Centric Design and Evaluation}
A key next step is to integrate user feedback systematically. While the current system was primarily evaluated through technical benchmarks, human-centered studies can reveal usability barriers, preferred answer formats, and the extent to which citation tracing enhances comprehension. Structured evaluations involving journalists, NGO workers, and policy analysts could measure perceived trustworthiness, cognitive effort saved, and the likelihood of adopting such tools in daily workflows.

\subsection{Scalability and Efficiency Improvements}
Scaling the system to process and serve thousands of long policy documents will require advances in both retrieval infrastructure and inference optimization. Techniques such as optimized HNSW index parameter tuning, distributed vector search, and model quantization can reduce latency while maintaining retrieval accuracy. These optimizations are especially critical for deployment in resource-constrained or offline environments.

\section{Conclusion}
\label{sec:conclusion}

This work presented \texttt{PolicyBot}, a retrieval-augmented generation pipeline designed explicitly for navigating dense, jargon-heavy policy documents. By combining semantic chunking, multilingual dense embeddings, multi-stage retrieval with reranking, and strict hallucination control mechanisms, the system delivers factually grounded answers with transparent source citations. The architecture prioritizes privacy and low-latency inference through local deployment, making it particularly suitable for under-resourced environments where cloud-based solutions are impractical.

The experimental evaluation systematically benchmarked critical components of the pipeline, including the choice of LLM, embedding model, and chunking strategy, using standardized datasets and relevance-based retrieval metrics. The resulting configuration—centered on the \texttt{Gemma 3n:e4b}, \texttt{gte-multilingual-base} embeddings, and semantic chunking—strikes a balance between accuracy, efficiency, and transparency.

Beyond technical contributions, this work highlights a broader principle: domain-specific RAG systems, when designed with traceability and verifiability as primary goals, can make complex governance information more accessible without compromising trust. The citation tracing interface not only provides factual grounding but also enables independent verification, empowering users to interpret and act upon policy information with greater confidence.

Looking ahead, the outlined future work opens a path toward a multilingual, legally aware, and user-validated \texttt{PolicyBot} that could serve as a dependable assistant for policy professionals, civil society organizations, and citizens worldwide. By maintaining a focus on transparency, fairness, and privacy, such systems can make a meaningful contribution to the democratization of policy understanding in an increasingly complex information landscape.

\section{Demo \& Source}
\label{sec:demo_source}

The source code of the \texttt{PolicyBot} is available on GitHub\footnote{https://github.com/cerai-iitm/policybot.git}. The instructions for using the demo are provided on the demo page. The datasets used in this study will be shared upon reasonable request.

\section*{Acknowledgments}
We thank the patrons of the Centre for Responsible AI\footnote{https://cerai.iitm.ac.in/} at IIT Madras for their continued support and assistance in building the PolicyBot.

\bibliographystyle{unsrt}  
\bibliography{references}

\end{document}